\documentclass[12pt]{article}

\newtheorem{theorem}{Theorem}
\newtheorem{corollary}{Corollary}
\newtheorem{lemma}{Lemma}

\usepackage{times,epsfig,graphics,amssymb,latexsym,amsmath,setspace,fullpage}
\usepackage{array,threeparttable,hyperref,url}
\usepackage[usenames]{color}

\def\argmin{\mathop{\rm argmin}}

\def\sign{\mathop{\rm sign}}

\begin{document}
\pagenumbering{arabic}

\setcounter{page}{1}

\title{\Large \bf On Minimum Clinically Important Difference}
\author{
 A. S. Hedayat, Junhui Wang and Tu Xu \\[10pt]
     Department of Mathematics, Statistics, \\
     and Computer Science \\
     University of Illinois at Chicago \\
     Chicago, IL, 60607 \\}
\date{}
\maketitle

\begin{abstract} \noindent In clinical trials, minimum clinically important difference (MCID) has attracted increasing interest as an important supportive clinical and statistical inference tool. Many estimation methods have been developed based on various intuitions, while little theoretical justification has been established. This paper proposes a new estimation framework of MCID using both diagnostic measurements and patient-reported outcomes (PRO's). The framework first formulates population-based MCID as a large margin classification problem, and then extends to personalized MCID to allow individualized thresholding value for patients whose clinical profiles may affect their PRO responses. More importantly, the proposed estimation framework is showed to be asymptotically consistent, and a finite-sample upper bound is established for its prediction accuracy compared against the ideal MCID. The advantage of our proposed method is also demonstrated in a variety of simulated experiments as well as two phase-3 clinical trials.
\end{abstract}


\footnotetext[1]{Research is supported in part by The U.S. National Science Foundation Grants DMS-0904125 (Hedayat) and DMS-1306394 (Hedayat).}
\footnotetext[2]{This article reflects the views of the author and should not be construed to represent FDA's view or policies.}

\doublespacing

\section{Introduction}
In clinical trials for drugs or medical devices, statistical
significance is widely used to infer the treatment effect. However, there has been growing recognition that statistical significance could be misleading when evaluating treatment effect (Jacobson et al., 1984; Jacobson and Truax, 1991). First, in many trials, the statistical significance of the treatment effect may have little to do with its clinical significance. It is known that statistical significance only infers the existence of treatment effect, regardless of the effect size. Further, the statistical significance could result from a small sample variability or a huge sample size, and thus provides little information about the clinical meaningfulness of the treatment (Jacobson and Truax, 1991). Second, the statistical significance for the treatment group compared to the placebo group ignores the possible heterogeneity among individuals. For instance, in a pain reduction study, a statistically significant reduction is concluded for a test treatment while many individual patients in the treatment group actually report little improvement regarding the pain reduction (Younger et al., 2009).

Clinical significance is desired in practice as it provides a better assessment of the clinically meaningful improvement. It is often based on the patients' reports in a community according to certain external standards (Jacobson and Truax, 1991). One common approach is to collect patient-reported outcomes (PRO's; FDA, 2009), such as their satisfaction of a treatment. Some earlier practice suggested to replace the statistical significance tests by analyzing the PRO's only, which is problematic due to the subjective bias in the PRO's or unreliability of a poorly designed questionnaire. Minimum clinically important difference (MCID) was discussed in Jaeschke et al. (1989), which was intuitively defined as a thresholding value in post-treatment change, and a patient is considered experiencing a clinically meaningful improvement if her/his change exceeds the MCID. Copay et al. (2007) suggested to incorporate both certainty of effective treatment and patients' satisfactions for determining MCID. 

The concept of MCID provides objective reference for clinicians and health policy makers regarding the effectiveness of the treatment, and has quickly gained its popularity among the practitioners. In 2012, U.S. Food and Drug Administration (FDA) hosted a special conference on the MCID for orthopaedic devices (\url{http://www.fda.gov/MedicalDevices/NewEvents/Workshops/Conferences/ucm327292.htm}). 
Although the importance of MCID has been widely recognized, only a few ad-hoc approaches have been proposed for its estimation with little theoretical justification (Bennett, 1985; Leisenring and Alonzo, 2000; Shiu and Gatsonis, 2008).

In this paper, the MCID is formulated as the thresholding value in post-treatment change such that the probability of disagreement between the estimated satisfaction based on the MCID and the PRO
 is minimized. With this framework, two scenarios are considered: population-based MCID and personalized MCID. The population-based MCID is the ideal thresholding value for the general population, and the personalized MCID allows different MCID values for individual patients based on their clinical profiles. Both scenarios can be formulated in a large margin classification framework, where the population-based MCID can be estimated via an exhaustive grid search, and the personalized MCID is modeled in a reproducing kernel Hilbert space and estimated via some non-convex optimization techniques. Most importantly, the asymptotic properties of the proposed estimation method are established for both population-based and personalized MCID's, and their fast convergence rates to the ideal performance are explicitly quantified.

The rest of the paper is organized as follows. In Section 2, a general framework for the population-based MCID is presented, and its estimation algorithm and asymptotic properties are studied.
Section 3 extends the framework to the personalized MCID, and discusses the appropriate large margin loss as well as the efficient non-convex optimization technique. Section 4 establishes the asymptotic properties of our proposed method for estimating the personalized MCID. Section 5 conducts numerical experiments of our proposed method in simulated examples, and Section 6 applies our proposed method to two phase-3 clinical trial datasets. Section 7 contains some discussion, and the appendix is devoted to technical proofs.

\section{A general framework of MCID}

\subsection{Formulating MCID}

Suppose that a patient's diagnostic measurement $X \in {\cal R}^1$ is continuously connected, and the patient-reported outcome (PRO) $Y \in \{-1,1\}$, where $Y = 1$ denotes a clinically meaningful treatment reported by the patient and $Y=-1$ otherwise.  Let $f(x,y)$ and $f(x)$ be the joint density of $(X,Y)$ and the marginal density of $X$, respectively. The MCID is formulated as the thresholding value $c^*$ such that $\sign(X - c^*)$ agrees with $Y$ as much as possible, where $\sign(u) = 1$ if $u \geq 0$ and $-1$ otherwise. Mathematically, $c^*$ is defined as a solution of 
\begin{equation}
\min_{c}~P(Y \neq \sign(X - c) )=\min_c~\frac{1}{2} E\left(1-Y \sign(X-c)\right),
\label{eqn:cmt}
\end{equation}
where $P(\cdot)$ is taken with respect to both $X$ and $Y$.

{\lemma
Assume that $p(x)=P(Y=1|X=x)$ is continuous and increasing in $x$, then the ideal MCID $c^*$ satisfies
\begin{equation}
p(c^*)=P(Y=1|X=c^*)=\frac{1}{2}.
\label{eqn:cmt_sol}
\end{equation}
Furthermore, if $p(x)$ is strictly increasing in $x$, then $c^*$ is the unique root of (\ref{eqn:cmt_sol}).
}

Note that it is reasonable to assume that $p(x)$ is increasing in $x$ since patients with better diagnostic measurements are expected to be more likely to give positive responses. If $p(x)$ is only non-decreasing, the equation in (\ref{eqn:cmt_sol}) may have multiple roots and a conservative choice is to set $c^*$ as the largest root. Furthermore, the continuity assumption of $p(x)$ can be relaxed to semi-continuity, and then the equation in (\ref{eqn:cmt_sol}) may have no root at all. In such scenarios, it could be proved similarly as Lemma 1 that $c^* = \argmin_c\{p(c) \ge 1/2\}$. 

It is known that the quality of the PRO's is largely affected by patients' subjectivity (Frost et al., 2007). Such subjectivity is accounted in the proposed formulation of the MCID through $p(x)$, which can be interpreted as the probability of patient's telling the truth. For instance, Fang (2011) considered a special case of semi-continuous $p(x)$, and modeled the subjectivity explicitly as $p(x)=Q$ when $x \geq c^*$ and $p(x)=1-Q$ otherwise, where $Q > \frac{1}{2}$ measures how trustworthy the PRO's are. More importantly, the ideal MCID in (\ref{eqn:cmt_sol}) is less affected by the subjectivity in the PRO's, as it relies on $p(x)$ only when $x$ is in the neighborhood of $c^*$. This is analogous to the Bayes rule in classification, which only relies on whether $p(x) \geq 1/2$ (Lin, 2002).

In addition, the MCID has an interesting connection with the median lethal dose in toxicology research. The median lethal dose refers to the smallest dose required to kill half of the animals that receive it after a specified test duration. To describe the interaction between dosage and mortality rate, the logistic dose-response curve is popularly used (Williams, 1986; Alho and Valtonen, 1995; Kelly, 2001) . It assumes that the mortality rate is expected to strictly increase with dose, which coincides with our assumption in Lemma 1.

\subsection{Estimating MCID}

The primary interest of this paper is to estimate the MCID, which is in sharp contrast to the standard classification that focuses on the classification boundary. In (\ref{eqn:cmt_sol}), the ideal MCID $c^*$ is defined based on $p(x)$ that is often unavailable in practice, so the MCID needs to be estimated based on the training sample $(x_i,y_i)_{i=1}^n$. 

Naturally, the expectation in (\ref{eqn:cmt}) can be approximated by its empirical version, and the estimated MCID $\hat{c}$ is defined as a solution of
\begin{equation}
\min_c~\frac{1}{2n} \sum_{i=1}^n \left(1-y_i \sign(x_i-c)\right).
\label{eqn:cmt_est}
\end{equation}
Note that (\ref{eqn:cmt_est}) is a simple 1-dimensional optimization problem, and the objective function remains the same for $x_{(i)} \leq c < x_{(i+1)}$, where $x_{(i)}$ is the $i$-th order statistic. Therefore, an exhaustive grid search scheme can be implemented, and the global minimizer $\hat c$ is simply the $x_i$ that yields the smallest objective function value.

{\theorem
 The estimated MCID $\hat c$ in (\ref{eqn:cmt_est}) is a consistent estimate of $c^*$ if  $p(x)$ is continuous and strictly increasing in $x$. Further, if there exist positive constants $\alpha_1$, $\gamma_1 < 2/\alpha_1 + 4/ \alpha_1^2$, $a_1$ and $a_2$, such that for sufficiently small $\xi>0$,
\begin{eqnarray}
P(|p(X) - p(c^*)| \leq \xi) \leq a_1 \xi^{\alpha_1},
\label{eqn:low_noise}  \\
\sup\limits_{|x-c^*|\leq \xi}|p(x)-p(c^*)|\leq a_2 \xi^{\gamma_1},
\label{eqn:lipschitz}
\end{eqnarray}
then $|\hat c - c^*| = O_p \Big((n \log^{-2}n)^{-1/(2(1+2/\alpha_1)-\alpha_1\gamma_1)}\Big)$. 
}

Theorem 1 establishes the asymptotic convergence rate of $|\hat c - c^*|$, and the finite sample bound for $|\hat c - c^*|$ can also be obtained as in Appendix A. In Theorem 1, (\ref{eqn:low_noise}) is similar to the low noise assumption (Polonik, 1995; Bartlett et al., 2003; Tsybakov, 2004) that describes the behavior of $X$ in the neighborhood of $c^*$, and (\ref{eqn:lipschitz}) is implied by a H\"{o}lder continuity condition on $p(x)$. For illustration, if $X$ is uniformly distributed on $[a,b]$ and (\ref{eqn:lipschitz}) is met with $\gamma_1$, then (\ref{eqn:low_noise}) can be verified with $\alpha_1 = 1/\gamma_1$ for sufficiently small $\xi$. Theorem 1 then implies that  $|{\hat c}-c^*|=O_p \big( (n \log^{-2} n)^{-1/(1+4\gamma_1)} \big)$. It leads to a fast convergence rate when $p(x)$ has a steep derivative at $c^*$ with $\gamma_1$ close to 0, and a rate of $O_p \big( n^{-1/3} (\log n)^{2/3} \big )$ when (\ref{eqn:lipschitz}) holds with order $\gamma_1=1/2$.

\subsection{Weighted MCID}

In many clinical studies, it is a common practice to be conservative when predicting whether the test outcome is clinically meaningful. It is then less desirable to predict positive for an unsatisfied patient than negative for a satisfied patient. To accommodate the unbalanced severity, the weighted MCID can be introduced with the weights reflecting the severity of the disagreements. Specifically, the weighted MCID $c_w^*$ is defined as a solution of 
\begin{equation}
\min_c~\frac{1}{2} E \Big ( w(Y) \left(1-Y \sign(X-c)\right) \Big ),
\label{eqn:cmt_w}
\end{equation}
where $w(1)=w$ and $w(-1)=1-w$. Similarly as in Lemma 1, it can be shown that
\begin{equation}
p(c_w^*)=P(Y=1|X=c_w^*)=1-w,
\label{eqn:cmt_w_sol}
\end{equation}
where an appropriate choice of $w<1/2$ leads to a conservative estimation.

The weighted MCID has another useful interpretation in the context of hypothesis testing. In particular, we denote the type-I error and type-II error as $R_{0}(c)=P(X-c > 0|Y=-1)$ and $R_{1}(c)=P(X-c < 0|Y=1)$, respectively. Then it is natural to find $c_\alpha^*$ to solve
\begin{equation}
\min\limits_{c} R_{1}(c)~~\mbox{subject to}~~R_{0}(c)\leq \alpha,
\label{eqn:neyman}
\end{equation}
where $\alpha$ is the significance level as in the standard hypothesis testing setup. This formulation is closely related with the Neyman-Pearson classification as discussed in Scott and Nowak (2005) and Rigollet and Tong (2011). More interestingly, an one-to-one correspondence between the weighted MCID $c_w^*$ in (\ref{eqn:cmt_w_sol}) and the solution $c_{\alpha}^*$ in (\ref{eqn:neyman}) can be easily established.

%More interestingly, Lemma 2 establishes a one-to-one correspondence between the weighted MCID $c_w^*$ in (\ref{eqn:cmt_w_sol}) and the solution $c_{\alpha}^*$ in (\ref{eqn:neyman}).

%{\lemma Assume that $p(x)$ is continuous and strictly increasing in $x$, then for any $\alpha$, there exists a unique $w$ such that  $c_\alpha^*=c_w^*$, and vice versa. }

%If $p(x)$ is only non-decreasing, there may exist multiple roots to the equation in (\ref{eqn:cmt_w_sol}) for certain $w$, whereas the solution $c^*_{\alpha}$ to (\ref{eqn:neyman}) is still uniquely defined for any given $\alpha$. In fact, it can be showed that $c^*_{\alpha}$ and $\alpha$ are one-to-one correspondent as long as $p(x)$ stays away from $0$ and $1$.

\section{Personalized MCID}

In many clinical trials, it is commonly believed that 
patients' report could be influenced by various factors such as their expectation of treatment (Wise, 2004). For instance, in a shoulder pain reduction study, healthy people demonstrate a higher threshold than those with chronic conditions due to their expectation of complete recovery. To allow the MCID to vary according to patients' clinical profiles, this section extends the estimation framework to personalized MCID.

\subsection{Formulation}

Let $z$ denote patients' clinical profiles, and the personalized MCID $c^*(z)$ is formulated as a solution of
\begin{equation}
\min_{c} P(Y \neq \sign(X - c(Z)) )=\min_{c} \frac{1}{2} E\left(1-Y \sign(X-c(Z))\right),
\label{eqn:cmt_z}
\end{equation}
where $P$ is taken with respect to $(X, Y, Z)$. Similarly as in (\ref{eqn:cmt_sol}), we can show that $c^*(z)$ satisfies
\begin{equation}
p_z(c^*(z))=P(Y=1|X=c^*(z),Z=z)=\frac{1}{2},
\label{eqn:cmt_z_sol}
\end{equation}
where $p_z(x)= P(Y=1|X=x,Z=z)$ is assumed to be a continuous and strictly increasing function in $x$ for any value of $z$. If only semi-continuity is assumed, the MCID can be formulated as $c^*(z)=\argmin_{c} \{c: p_z(c) \geq \frac{1}{2}\}$. It is worth pointing out that the personalized MCID in (\ref{eqn:cmt_z}) differs from the Bayes rule in classification in that the candidate function in (\ref{eqn:cmt_z}) has to take the form of $x-c(z)$ in order to estimate $c^*(z)$, whereas a Bayes rule in classification searches for the optimal classification function $g(x,z)$ that may not lead to an explicit estimation of $c^*(z)$.

The formulation in (\ref{eqn:cmt_z}) is similar as in (\ref{eqn:cmt}) with population-based $c^*$, but the difficulty arises in the estimation part. Since the empirical version of (\ref{eqn:cmt_z})
\begin{equation}
\min_{c} \frac{1}{2n} \sum_{i=1}^n \Big(1-y_i \sign(x_i-c(z_i))\Big),
\label{eqn:cmt_z_est}
\end{equation}
involves the 0-1 loss $L_{01} (u)=\frac{1}{2}(1-\sign(u))$ and needs to be optimized with respect to functional $c(z)$, it can no longer be solved by the exhaustive grid search or any other efficient optimization techniques. Therefore, a surrogate loss function needs to be introduced to replace the 0-1 loss and facilitate the estimation. The surrogate loss has been widely studied in machine learning literature. Popularly used surrogate loss functions $L(u)$ include the hinge loss $L(u)=(1-u)_+$ (Vapnik, 1998), the logistic loss $L(u)=\log(1+\exp(-u))$ (Zhu and Hastie, 2005), and the $\psi$-loss $\min((1-u)_+,1)$ (Shen et al., 2003; Liu and Shen, 2006). However, all these losses are not generally Fisher consistent in estimating $c^*(z)$, as the candidate function in (\ref{eqn:cmt_z}) is restricted to the form of $x-c(z)$ for estimating MCID. Counter examples can be constructed as in Appendix B.

In this paper, we propose a novel surrogate loss, $\psi_{\delta}$-loss, which is defined as
\begin{equation}
L_{\delta} (u)=\min \left( \frac{1}{\delta}(\delta-u)_+,1 \right ).
\label{eqn:psi_loss}
\end{equation}
The $\psi_{\delta}$-loss extends the $\psi$-loss by introducing a new parameter $\delta$ that controls the difference between the surrogate loss and the 0-1 loss. More importantly, Lemma 2 shows that the $\psi_{\delta}$-loss is asymptotically Fisher consistent in estimating $c^*(z)$ when $\delta$ converges to 0.

{\lemma For any given $z$, if the conditional density $f_z(x)$ is continuous and $p_z(x)$ is strictly increasing in $x$, then $E \Big ( L_{\delta}( Y(X-c))|Z=z \Big )$ converges to $E \Big ( L_{01}( Y(X-c) )|Z=z \Big )$ as $\delta \rightarrow 0 $ uniformly over a compact set ${\cal D}_z$ containing $c^*(z)$ and
$$
\argmin_{c} E \Big ( L_{\delta}( Y(X-c(z)) ) |Z=z\Big ) \longrightarrow c^*(z).
$$}

With the $\psi_{\delta}$-loss, the proposed estimation formulation for the personalized MCID $\hat c(z)$ solves
\begin{equation}
\min_{c \in {\cal F}} \frac{1}{n} \sum_{i=1}^n L_{\delta}( y_i(x_i-c(z_i)) ) + \lambda J(c),
\label{eqn:cmt_z_cost}
\end{equation}
where $\lambda$ is a tuning parameter, $J(c)$ is a penalty term, and $\cal F$ is set as a reproducing kernel Hilbert space (RKHS; Wahba, 1990).
%\change{Reproducing kernel Hilbert spaces (RKHS; Wahba, 1990) are popularly used for estimating complex functions due to its theroetical and comptuational advantages (Shen et al., 2003; Zhu and Hastie, 2005; Li et al., 2007). In this paper, we set $\cal F$ as a RKHS,} 
The final estimation formulation then becomes
\begin{equation}
\min_{c \in {\cal H}_K}~\frac{1}{n} \sum_{i=1}^n  L_{\delta}( y_i(x_i-c(z_i)) )  + \frac{\lambda}{2} \|c\|^2_{{\cal H}_K},
\label{eqn:rkhs}
\end{equation}
where ${\cal H}_K$ is the RKHS induced by some pre-specified kernel function $K(\cdot,\cdot)$, and $J(c)=\frac{1}{2}\| c
\|_{{\cal H}_K}^2$ is the associated RKHS norm of $c(z)$. It follows from the representer theorem (Wahba, 1990) that the solution to (\ref{eqn:rkhs}) is of the form $\hat c(z) = b + \sum_{i=1}^n w_i K(z_i,z)$, and thus $\| c
\|_{{\cal H}_K}^2=w^T {\bf K} w$ with $w=(w_1,\cdots,w_n)^T$ and ${\bf K}=(K(z_i,z_j))_{i,j=1}^n$.

\subsection{Non-convex optimization}

Note that the cost function in (\ref{eqn:rkhs}) is non-convex, and thus we employ the difference convex algorithm (DCA; An and Tao, 1997) to tackle the non-convex optimization. The key idea of the DCA is to decompose the non-convex cost function into the difference of two convex functions, and then construct a sequence of subproblems by approximating the second convex function with its affine minorization function. 

In particular, the $\psi_{\delta}$-loss is decomposed as 
$$
L_{\delta} (u)=\min \left( \frac{1}{\delta}(\delta-u)_+,1 \right )=\frac{1}{\delta}(\delta-u)_+ - \frac{1}{\delta}(-u)_+.
$$
Then the cost function in (\ref{eqn:rkhs}) can be decomposed as $s(w)=s_1(w)-s_2(w)$, where
\begin{eqnarray*}
s(w) &=& \frac{1}{n} \sum_{i=1}^n L_{\delta}( y_i(x_i-c(z_i)) ) + \frac{\lambda}{2} \|c\|^2_{{\cal H}_K}, \\
s_1(w) & = &\frac{1}{n} \sum_{i=1}^n  \Big (\frac{1}{\delta}\left(\delta-y_i(x_i-c(z_i))\right)_+\Big ) + \frac{\lambda}{2}\| c \|_{{\cal H}_K}^2, \\
s_2(w) & = &\frac{1}{n} \sum_{i=1}^n  \Big (\frac{1}{\delta}\left(-y_i(x_i-c(z_i))\right)_+\Big ),
\end{eqnarray*}
and $w$ is the coefficient vector for the RKHS representation of $c(z)$. 

Next, the DCA constructs a sequence of decreasing upper envelop of $s(w)$ by approximating $s_2(w)$ with its affine minorization function, $s_2(w^{(k)})+\langle w-w^{(k)}, \nabla s_2 (w^{(k)})\rangle$, where $w^{(k)}$ is the estimated $w$ at the $k$-th iteration, and $\nabla s_2 (w^{(k)})$ is the subgradient of $s_2(w)$ at $w^{(k)}$. The updated $w^{(k+1)}$ is then obtained by solving
\begin{equation}
w^{(k+1)}= \argmin_w~s_1(w)-s_2(w^{(k)})-\langle w-w^{(k)}, \nabla s_2 (w^{(k)})\rangle.
\label{eqn:DCA}
\end{equation}
The updating scheme is iterated until convergence. Although the DCA cannot guarantee global optimum, it delivers a superior numerical performance as demonstrated in the extensive simulation study in Liu et al. (2005).

\section{Asymptotic theory}

This section quantifies the asymptotic behavior of $\hat c(z)$ in estimating the personalized MCID. 
%The estimation accuracy of $\hat c(z)$ is measured by $|\hat c(Z) - c^*(Z)|$. 
%$$
%e(\hat c, c^*)=E \Big( L_{01}(Y(X-\hat c(Z))) - L_{01}(Y(X-c^*(Z))) \Big).
%$$
Denote $e_{\delta_n}(\hat c, c^*)=E \Big( L_{\delta_n}(Y(X-\hat c(Z))) - L_{\delta_n}(Y(X-c^*(Z))) \Big)$ with $\delta_n >0$, where $\delta$ and $\lambda$ are rewritten as $\delta_n$ and $\lambda_n$ to denote their dependency on $n$. We make the following four technical assumptions.

{\it Assumption A.} For some positive sequence $s_n \rightarrow 0$ as $n \rightarrow \infty$, there exists $c_0(z) \in {\cal F}$, such that for sufficiently small $\delta_n$, $e_{\delta_n}(c_0, c^*) \leq s_n$. That is, $\inf_{\{c \in {\cal F}\}} e_{\delta_n} (c,c^*) \leq s_n$.

Assumption A is standard (Shen et al., 2003; Li et al., 2007), and describes the approximation error of $\cal F$ in approximating $c^*(z)$.

{\it Assumption B.} There exist constants $0 <\alpha_2 < +\infty$ and $a_3 >0$ such that for any given $z$, $P ( |p_z(X)-p_z(c^*(z))|\leq \xi) \leq a_3 \xi^{\alpha_2}$ for sufficiently small $\xi$.

Assumption B is the low noise assumption that describes the distribution of the diagnostic outcome $X$ in the neighborhood of $c^*(z)$.

{\it Assumption C.} There exist constants $0 <\gamma_2 < +\infty$ and $a_4 >0$ such that for any given $z$, $\sup_{|x-c^*(z)|\leq \xi}|p_z(x)-p_z(c^*(z))| \leq a_4 \xi^{\gamma_2}$ for sufficiently small $\xi$. 

Assumption C is implied by a H\"{o}lder continuity condition that describes the smoothness of $p_z(x)$ around $c^*(z)$.

Before specifying Assumption D, we first define the metric entropy for any give set. For a given class ${\cal B}$ of subsets of $S$ and any $\epsilon >0$, $\{(G_1^l, G_1^u, \cdots, G_m^l, G_m^u)\}$ forms an $\epsilon$-bracketing set of ${\cal B}$ if for any $G \in {\cal B}$ there is a $j$ such that $G_j^l \subset G \subset G_j^u$ and $\max_{\{1\leq j \leq m\}}d(G_j^u, G_j^l) \leq \epsilon$, where $d(\cdot,\cdot)$ is a distance for any two subsets in $S$ defined as $d(G_1, G_2) = P(G_1 \Delta G_2)$ and $G_1 \Delta G_2 = (G_1 \backslash G_2) \bigcup (G_2 \backslash G_1)$. Then the metric entropy $H(\epsilon, {\cal B})$ of ${\cal B}$ is defined as the logarithm of the cardinality of the $\epsilon$-bracketing set of ${\cal B}$ of the smallest size. Let ${\cal G}(k)=\{G_c=\{(x,z):x-c(z)\geq 0\},c \in {\cal F}, J(c) \leq k\} \subset {\cal G}({\cal F})=\{G_c=\{(x,z):x-c(z)\geq 0\},c \in {\cal F}, J(c) < +\infty\}$.

{\it Assumption D.} For positive constants $a_5$, $a_6$ and $a_7$, there exists some $\epsilon_n>0$ such that
$$\sup\limits_{\{k \geq 1\}} \phi(\epsilon_n,k)\leq a_5 n^{1/2},$$
where $\phi(\epsilon_n,k)= \int_{a_7 L}^{(8a_6)^{1/2} L^{\alpha/2(\alpha+\gamma)}}H^{1/2}(u^2/2, {\cal G}(k)) du/L$ and $L=L(\epsilon_n, C, k)=\min(\epsilon_n^2 + \lambda_n J_0(k/2-1),1)$.

{\theorem Suppose that Assumptions A-D are met. For the estimated personalized MCID $\hat{c}(z)$, there exists positive constants $a_{8}$ and $a_9$ such that
$$
P \Big (|\hat c(Z)-c^*(Z)| \geq (\beta_n^2 \log(1/\beta_n^2) )^{\frac{ \alpha_2}{\alpha_2+2}}  \Big )  \leq  3.5 \exp \Big ( -a_{8} n (\lambda_n J(c_0))^{\frac{\alpha_2+2}{\alpha_2+1}} \Big ) + a_9 (\log(1/\beta_n^2))^{-1}.
$$
provided that $\beta_n^2 \geq 4\lambda_n \max(J(c_0),1)$ with $\beta_n^2 = \min(\max(\epsilon_n^2, 2s_n+2a_3 a_4^{\alpha_2} \delta_n^{\alpha_2 \gamma_2}),1)$ and $f_z(c^*(z))$ is bounded away from 0.
}

{\corollary Under the assumptions of Theorem 2, $|\hat c(Z)-c^*(Z)| =O_p((\beta_n^2 \log(1/\beta_n^2) )^{\frac{ \alpha_2}{\alpha_2+2}}) $,
%E|\hat c(Z)-c^*(Z)|=O((\beta_n^2 \log(1/\beta_n^2) )^{\frac{ \alpha_2}{\alpha_2+2}}),
%$$
provided that $n (\lambda_n J(c_0))^{\frac{\alpha_2+2}{\alpha_2+1}}$ is bounded away from 0.
}

Theorem 2 and Corollary 1 develop upper bounds for the estimation accuracy of the estimated $\hat c(z)$. The convergence rate $\beta_n^{\frac{2 \alpha_2}{\alpha_2+2}}$ in Corollary 1 depends on the value of $\delta_n$, $\epsilon_n^2$, $s_n$ and $\lambda_n$. 
%In particular, when $\delta_n=O(s_n^{1/(\alpha_2\gamma_2)})$, the convergence rate becomes $O_p(\min(\max(\epsilon_n^2, 2s_n),1))$. 
%More importantly, Corollary 1 establishes asymptotic convergence rate for the estimation of the personalized MCID $c^*(z)$. 
More importantly, such results can be difficult to establish for the standard classification function $g(x,z)$ due to its lack of explicit estimation of $c^*(z)$.

\section{Simulation}

This section examines the proposed estimation methods for estimating MCID using simulated examples. Two scenarios are considered. Scenario I focuses on the population-based MCID for all patients, and scenario II focuses on the personalized MCID that varies among patients and relies on each patient clinical profile. To assess the estimation performance, we report the estimated MCID as well as the misclassification error (MCE) based on the testing set, which is defined as 
$$
MCE(\hat c)= \frac{1}{n_{test}} \sum_{i \in testing~set} I(y_i \neq \sign(x_i-\hat c(z_i))),
$$
where $n_{test}$ denotes the size of the testing set, and $\hat c(z_i)=\hat c$ for the population-based MCID.

\subsection{Scenario I: population-based MCID}

Two simulated examples are examined.

{\it Example 1.} A random sample $\{(X_i,Y_i); i=1,\cdots, n+2000\}$ is generated as follows. First, $X_i$ is generated from $Unif(-1,1)$ and then $Y_i$ is generated from $Bern((x_i+1)/2)$. Next, a sample of size $n$ is randomly selected for training and the remaining $2000$ samples are allocated for testing.

{\it Example 2.} A random sample $\{(X_i,Y_i); i=1,\ldots, n+2000\}$ is generated as follows. First, $X_i$ is generated from the mixture of two Gaussian distributions $0.7 N(-1,1) + 0.3 N(1,1)$ and then $Y_i$ is generated from $Bern(F(x_i))$, where $F(x_i)= P(X \leq x_i)$. Next, a sample of size $n$ is randomly selected for training and the remaining $2000$ samples are allocated for testing.

In both examples, the training sizes are set as $n=250$, 500 and 1000. Both examples are replicated 100 times. The averaged performance measures of our proposed method and Shiu and Gatsonis (2008) are reported in Table \ref{tab:sim1}. In addition, the ideal MCID's and their corresponding misclassification errors are used as baseline for the comparison in Table \ref{tab:sim1}.

\begin{center}
\begin{tabular}[t]{c}
      \hline
      \hline
      Table \ref{tab:sim1} about here\\
      \hline
      \hline
\end{tabular}
\end{center}

In both examples, our proposed method yields accurate MCID estimates that are very close to the ideal MCID's. The resulting MCE's are also close to the MCE's produced by using the ideal MCID's. The performance of the method by Shiu and Gatsonis appears to be less competitive. Even with a large sample size $n=1000$, their estimated MCID's are still considerably different from the ideal MCID's.

\subsection{Scenario II: personalized MCID}

For personalized MCID, the MCE by using our proposed method with linear and Gaussian kernels are examined. The linear kernel is defined as $K(z_1,z_2)=z_1^T z_2$, and the Gaussian kernel is defined as $K(z_1,z_2)=e^{-\|z_1-z_2\|^2/2 \sigma^2}$, where the scale parameter $\sigma^2$ is set as the median of pairwise Euclidean distances within the training set. To optimize the performance of our proposed method, a grid search by 5-fold cross validation is employed to select the tuning parameter $\lambda$. The grid for all examples is set as $\{10^{(s-31)/10}; s=1,\cdots,61\}$. For illustration, three simulated examples are examined with $\delta=0.1$. 

{\it Example 1.} A random sample $\{(X_i,Y_i,Z_i); i=1,\cdots, n\}$ is generated as follows. First, $Z_i$'s are independently generated from $N_2(\mu,I_2)$ with $\mu=(0,0)^T$. Second, $X_i$'s are independently generated from $N(b+w^T z_i,1)$, where $b=0$ and $w=(1,2)^T$. Next, the response $Y_i$ is generated from $Bern(F(x_i))$, where $F(x_i)=P(X_i \leq x_i)$. 

{\it Example 2.} A random sample $\{(X_i,Y_i,Z_i); i=1,\cdots, n\}$ is generated as follows. First, $Z_i$'s are independently generated from $N_2(\mu,I_2)$ with $\mu=(0,0)^T$. Second, $X_i$'s are independently generated from $N(b+w^T z_i-w^T z_i^2,1)$, where $b=0$ and $w=(1,2)^T$. Next, the response $Y_i$ is generated from $Bern(F(x_i))$, where $F(x_i)=P(X_i \leq x_i)$.

{\it Example 3.} A random sample $\{(X_i,Y_i,Z_i); i=1,\cdots, n\}$ is generated as follows. First, $Z_i$'s are independently generated from $N_3(\mu,I_3))$ with $\mu=(0,0,0)^T$. Second, $X_i$'s are independently generated from $N(b+\cos(w^T z_i),1)$, where $b=0$ and $w=(1,1.5,2)^T$. Next, the response $Y_i$ is generated from $Bern(F(x_i))$, where $F(x_i)=P(X_i \leq x_i)$. 

For each example, the training sizes are set as 100, 250, 500 and the testing size is set as 2000. All examples are replicated 50 times, and the averaged test errors are reported in Table \ref{table2}.

\begin{center}
\begin{tabular}[t]{c}
      \hline
      \hline
      Table \ref{table2} here\\
      \hline
      \hline
\end{tabular}
\end{center}

Our proposed method delivers satisfactory performance in estimating the personalized MCID in all three examples. In addition, the linear kernel yields slightly better performance than the Gaussian kernel in Example 1 as the true classification boundary is linear,
%(Figure \ref{fig:EstC}), 
and it is outperformed by the Gaussian kernel in the other two examples with nonlinear boundaries. Therefore, the Gaussian kernel would be suggested if no prior knowledge about the boundary is available.

%\begin{center}
%\begin{tabular}[t]{c}
%      \hline
%     \hline
%      Figure \ref{fig:EstC} about here\\
%     \hline
%     \hline
%\end{tabular}
%\end{center}
 
For estimating the personalized MCID, the choice of $\delta$ may impact the performance of our proposed method. By Theorem 2, large $\delta$ leads to less accurate prediction while computational instability may occur when small $\delta$ is used for the estimation. For illustration, we conducted a sensitivity analysis on the values of $\delta$ in a random replication of Example 1 with training size 250. The estimated coefficients and prediction error as functions of $\delta$ are displayed in Figure 1. It is evident that when $\delta$ is too large,  the estimation of $c(z)$ moves away from the truth and yields a larger error rate. When $\delta$ is close to 0, the error rate and estimation of $c(z)$ are relatively stable. Therefore, we recommend to set $\delta$ as 0.1 for simplicity.       

\begin{center}
\begin{tabular}[t]{c}
      \hline
      \hline
      Figure \ref{fig:Sensitivity} about here\\
      \hline
      \hline
\end{tabular}
\end{center}

\section{Real applications}

In this section, our proposed method is applied to a phase-3 woman heavy menstrual blood loss dataset (WHMBL) and a phase-3 hot flush dataset (Hot Flush). 

The WHMBL clinical trial aims to develop a treatment for reducing the amount of blood loss during a menstrual cycle in excessive bleeding women. The primary efficacy variable is the change from baseline in blood loss volume. The blood loss of each patient is measured per menstrual cycle and the PRO's are collected based on a questionnaire answered by each patient at a post-treatment visit. The WHMBL trial dataset consists of 481 patients administered either placebo or active doses. Patient profile contains the information of age, body mass index (BMI), alcohol (Yes/No), tobacco (Yes/No) and baseline value of blood loss. The 481 patients were randomly split into a training set of 240 patients and a testing set of 241 patients. 

The hot flush clinical trial aims to develop a treatment for reducing hot flush in women due to menopause. The hot flush clinical trial dataset consists of 1684 patients administered either placebo or active doses. Patient profile contains the information for age, BMI, race and baseline hot flushes. 300 patients were selected randomly to form the training set and the remaining 1384 patients were used as the testing set.

Here, $\delta=0.1$ is used for simplicity and the tuning parameter $\lambda$ is selected as in Section 5.2. Each example is replicated 50 times, and Table 4 summarizes the averaged performance measures of the method by Shiu and Gatsonis, the population-based MCID, and the personalized MCID with the linear and Gaussian kernels. 

\begin{center}
\begin{tabular}[t]{c}
      \hline
      \hline
      Table \ref{table4} about here\\
      \hline
      \hline
\end{tabular}
\end{center}

In both scenarios, our proposed method delivers competitive performance in comparison with the method by Shiu and Gatsonis. In WHMBL trial, the method by Shiu and Gatsonis yields a negative MCID which is clinically misleading. It is also interesting to notice that for the WHMBL trial, personalized MCID yields larger MCE when compared with population-based MCID. It could be due to the homogeneity among the enrolled patients. For the hot flush trial, patients' satisfaction on treatment effect is more accurately estimated when the clinical profiles are included. A closer investigation of the fitted classification function implies that patients' satisfaction is highly affected by the baseline hot flushes. This is reasonable as patients with higher baseline hot flushes tend to expect better treatment effect.

\section{Closing remarks}

This paper proposes a general framework for formulating as well as estimating population-based and personalized MCID's. The concept of MCID has attracted much attention in clinical trials, while little statistical work has been done for appropriately determining MCID. Our proposed method unifies both population-based and personalized MCID's into a large margin classification framework, and delivers superior estimation performance in both simulated examples and real applications to two phase-3 clinical trials. More importantly, the asymptotic properties of our proposed method are established for both population-based and personalized MCID's. Future research work will focus on the potential issues when applying our proposed MCID's to various clinical trials.

\section*{Acknowledgment}

The authors gratefully acknowledge the ORISE internship program supported by the Office of Biostatistics, CDER of FDA. In particular, the authors wish to acknowledge Dr. Xin Fang's contribution on this project, and thank Dr. Stephen E. Wilson, Dr. Ram Tiwari and Dr. Lisa LaVange for their strong support of this project and its associated internship program.

\section*{Appendix A: technical proofs}

{\bf Proof of Lemma 1.} Note that $c^*$ is a solution of
$$
\min_c \frac{1}{2} E\big(1-Y \sign(X-c)\big)=\min_c \frac{1}{2} E_X\big(1-E(Y|X) \sign(X-c)\big),
$$
where $E_X$ represents the expectation with respect to $X$. It then suffices to find $c^*$ to maximize $E(Y|X=x) \sign(x-c)$ for any given $x$. Therefore, $c^*$ must satisfy that
\begin{equation}
\sign(x-c)=\sign(E(Y|X=x))=\sign(2p(x)-1),
\label{eqn:lemma1}
\end{equation}
for any $x$, where $p(x)=P(Y=1|X=x)$. We now show contradiction when $p(c^*) \neq 1/2$. Without loss of generality, assume $p(c^*)>1/2$. Since $p(x)$ is continuous and monotone in $x$, there must exist $\tilde c$ such that $p(\tilde c)=1/2$ and $\tilde c <c^*$. This leads to the contradiction to (\ref{eqn:lemma1}) since
$$
0>\sign(\tilde c-c^*)=\sign(2p(\tilde c)-1)=1.
$$
Therefore, $c^*$ must satisfies $p(c^*)=\frac{1}{2}$. Furtheremore, when $p(x)$ is continuous and strictly increasing, the uniqueness follows from the fact that $p(c^*)=\frac{1}{2}$ has a unique solution.

%%%%%%%%%%%%%%%%%%%%%%%%%%%%%%%%%%%%%%%%%%%%%%%%%%%%%%%%%%%%%%%%%%%%%%%%%%%%%%%%%%%%%%%%%%%%%%%%%%%%%%%%%%%%%%%%%%%%
\noindent {\bf Proof of Theorem 1.} We first show that $\hat{c} \stackrel{p}{\longrightarrow} c^*$. Let $F_y(x)=P(X \le x, Y=y)$, then
\begin{eqnarray*}
\frac{1}{2} E\left(1-Y \sign (X-c)\right) &=& P(X \leq c, Y=1)+P(X > c,Y=-1)\\
& =  & F_1(c)+P(Y=-1)-F_{-1}(c).
\end{eqnarray*}
By strong law of large number, $\frac{1}{n} \sum_{i=1}^n I(Y_i=-1) \stackrel{a.s.} {\longrightarrow} P(Y=-1)$. Further, it follows from Theorem 19.1 of Van der Vaart (1998) that
%the empirical distribution function $F_n(\cdot)$ uniformly almost surely convergent to its distribution function $F(\cdot)$. Thus
\begin{eqnarray*}
F_{1,n}(c) &=& \frac{1}{n} \sum_{i=1}^n I(X_i \le c,Y_i=1) \stackrel{a.s.} {\longrightarrow} F_1(c), \\
F_{-1,n}(c) &=& \frac{1}{n} \sum_{i=1}^n I(X_i \le c,Y_i=-1) \stackrel{a.s.} {\longrightarrow} F_{-1}(c),
\end{eqnarray*}
uniformly over $c$. Therefore,
$$
\frac{1}{2n} \sum_{i=1}^n \left(1-y_i \sign(x_i-c)\right) \stackrel{a.s.} {\longrightarrow} \frac{1}{2} E\left(1-Y \sign(X-c)\right)
$$
uniformly over $c$. Also by Lemma 1, $\frac{1}{2} E\left(1-Y \sign(X-c)\right)$ has a unique minimizer $c^*$ when $p(x)$ is continuous and strictly increasing in $x$. The desired asymptotic consistency follows immediately after Theorem 5.7 of Van der Vaart (1998).

Next, we establish the convergence rate of $|\hat c -c^*|$ by using Theorem 5.52 of Van der Vaart (1998). We just need to verify the necessary assumptions. Note that $c^*$ is the minimizer of $\frac{1}{2}E(1-y \sign(x-c))$. Without loss of generality, for any $c > c^*$, direct deviation yields that
\begin{eqnarray*}
E(m_c(X,Y)-m_{c^*}(X,Y))  &=& P(c^* \leq X < c, Y=1)-P(c^* \leq X < c, Y=-1 ) \\
&=& \int_{c^*}^{c} p(x)f(x)dx - \int_{c^*}^{c}    (1-p(x))f(x)dx \\
&=& \int_{c^*}^{c} (2p(x)-1)f(x)dx,
\end{eqnarray*}
where $m_c(x,y) = \frac{1}{2}(1-y \sign(x-c))$.  

Since $f(x)$ is continuous at $c^*$, it can be shown that $ P(c^* \leq X \leq c^*+\xi) \geq  a_9 \xi$ for sufficient small $\xi>0$, where $a_{10}=f(c^*)/2>0$.  Furthermore, $ p(c^*+\xi)-p(c^*) > (a_{10}/a_1)^{1/\alpha_1}(\xi)^{2/\alpha_1}$, since otherwise there exists $0<\xi_1<1$ such that $p(c^*+\xi_1)-p(c^*) \leq (a_{10}/a_1)^{1/\alpha_1} (\xi_1)^{2/\alpha_1}$, and by assumption (\ref{eqn:low_noise})
$$
a_{10} \xi_1 \leq  P(c^* \leq X \leq c^*+\xi_1) \leq P \Big (|p(X) - p(c^*)| \leq (a_{10}/a_1)^{1/\alpha_1}(\xi_1)^{2/\alpha_1} \Big ) \leq a_{10} (\xi_1)^{2},
$$
which leads to a contradiction to the fact that $\xi_1<1$. 

Since $p(x)$ in continuous in $x$, there exists $0 < \xi_2 < \xi$ such that $p(c^*+\xi_2)-p(c^*)= (a_{10}/a_1)^{1/\alpha_1}(\xi)^{2/\alpha_1}$, and then
%for any sufficiently small $\xi>0$, let $\xi_1$ satisfy $p(\xi_1)-p(c^*)=\xi^{\gamma_1}$. Since $p(x)$ is increasing, there is $c^* < \xi_1 < \xi$. Then 
\begin{eqnarray*}
& & E(m_{c^*+\xi}(X,Y)-m_{c^*}(X,Y)) \\
&=& \int_{c^*}^{c^*+\xi} (2p(x)-1)f(x)dx > \int_{c^*+\xi_2}^{c^*+\xi} (2p(x)-1)f(x)dx >  (a_{10}/a_1)^{1/\alpha_1}(\xi)^{2/\alpha_1} \int_{c^*+\xi_2}^{c^*+\xi} f(x)dx \\
&=& (a_{10}/a_1)^{1/\alpha_1}(\xi)^{2/\alpha_1} \big(P(c^* \leq X \leq c^*+\xi)-P(c^* \leq X \leq c^*+\xi_2)\big)\\
&\geq& (a_{10}/a_1)^{1/\alpha_1}(\xi)^{2/\alpha_1} \Big(P(c^* \leq X \leq c^*+\xi)- P(|p(X)-p(c^*)| \leq (a_{10}/a_1)^{1/\alpha_1}(\xi)^{2/\alpha_1})\Big)\\
&\geq& a_{10}^{1+1/\alpha_1}a_1^{-1/\alpha_1}\xi^{2/\alpha_1} ( \xi-  \xi^2).
\end{eqnarray*}
It can be shown similarly that 
$$
E(m_{c^*-\xi}(X,Y)-m_{c^*}(X,Y)) \geq a_{10}^{1+1/\alpha_1}a_1^{-1/\alpha_1}\xi^{2/\alpha_1} ( \xi-  \xi^2).
$$ 
Therefore, there exists constant $a_{11}>0$ such that for sufficiently small $\xi>0$, 
\begin{equation}
\label{theorem1_1}
\sup\limits_{|c-c^*| < \xi} E(m_c(X,Y)-m_{c^*}(X,Y)) \geq a_{11} \xi^{1+2/\alpha_1}.
\end{equation}

Furthermore, denote ${\cal F}_m=\{m_c(x,y)-m_{c^*}(x,y):x \in {\cal R},~y\in \{-1,+1\}\}$. Consider the grid $-\infty=t_1 < t_1 < \cdots < t_k =+\infty$ with $t_{\left\lceil k/2 \right\rceil}=c^*$ and $P(x < t_i)-P(x \leq t_{i-1}) < \epsilon$ for each $t_i$. Note that
\begin{eqnarray*}
m_c(x,y)-m_{c^*}(x,y)= \left \{ 
\begin{array}{ll}
I(c^* \leq x < c, y=-1) - I(c^* \leq x < c, y=1),& \mbox{if}~c>c^*, \\
I(c < x \leq c^*, y=1) - I(c < x \leq c^*, y=-1),& \mbox{if}~c \leq c^*.
\end{array}
\right .
\end{eqnarray*}
Then the functional brackets $[1_{[c^*,t_{i}]}(x),1_{[c^*,t_{i+1})}(x)]$ for $i > \left\lceil k/2 \right\rceil$ and $[1_{[t_{i},c^*]}(x),1_{(t_{i-1},c^*]}(x)]$ for $i \leq \left\lceil k/2 \right\rceil$ forms $L_1(P)$ brackets of size $\epsilon$ for ${\cal F}_m$ with cardinality $k < 2/\epsilon$. Thus
the bracketing number $N_{[\ ]}(\epsilon, {\cal F}_m, L_2(P))=O(\epsilon^{-2})$ and then the bracketing integral 
$$
J_{[\ ]}(\eta, {\cal F}_m, L_2(P)) = \int_0^\eta \sqrt{\log N_{[\ ]}(\epsilon, {\cal F}_m, L_2(P))} d \epsilon \leq  a_{12} \eta \log \eta,
$$ 
for some constant $a_{12} < 0$.

Also $g(x) = I(c^*-\xi \leq x \leq c^*+\xi)$ is an envelop function of $m_c-m_{c^*}$ with $|c-c^*|<\xi$, and then assumptions (\ref{eqn:low_noise}) and (\ref{eqn:lipschitz}) imply that 
$$
\|g\|_{P,2}=(P(|X - c^*|\leq \xi))^{1/2} \leq \big(P(|p(X) - p(c^*)| \leq a_2 \xi^{\gamma_1})\big)^{1/2} \leq (a_1 a_2^{\alpha_1})^{1/2} \xi^{\alpha_1 \gamma_1/2} .
$$ 
By Corollary 19.35 of  Van der Vaart (1998),
\begin{eqnarray*}
E^* \sup\limits_{|c-c^*|< \xi} |G_n(m_c - m_{c^*})| &\leq&  J_{[\ ]}(\|g\|_{P,2}, {\cal F}_m, L_2(P)) \leq J_{[\ ]}( (a_1 a_2^{\alpha_1})^{1/2} \xi^{\alpha_1 \gamma_1} , {\cal F}_m, L_2(P)) \\
&\leq&   \frac{1}{2} a_{12} \alpha_1 \gamma_1 (a_1 a_2^{\alpha_1})^{1/2} \xi^{\alpha_1 \gamma_1/2}\log \xi.
\end{eqnarray*} 
Thereupon, denote $A=1+2/\alpha_1-\alpha_1 \gamma_1/2$, it follows from Theorem 5.52 of Van der Vaart (1998) that for a sufficiently large integer $M$,
\begin{eqnarray}
P^*\Big(|\hat{c} - c^*| \geq 2^M (n \log^{-2} n)^{-1/(2 A)}\Big) 
& \leq & \frac{2^{1+2/\alpha_1}}{ a_{11} A (2^{-A}-1)} a_{12} \alpha_1 \gamma_1 (a_1 a_2^{\alpha_1})^{1/2}  2^{-M A }.  \label{eqn:Theorem1}
\end{eqnarray}
%Thus, for $n$ satisfies $2^M \leq \log n < 2^{M+1}$, there is 
%$$
%P^*\Big(|\hat{c} - c^*| > \log n (n \log^{-2}n)^{-1/(2(1+2/\alpha_1)-\alpha_1 \gamma_1)}\Big) \leq 
%\frac{2^{4/\alpha_1-\alpha_1 \gamma_1/2+3}}{2^{1+2/\alpha_1-\alpha_1 \gamma_1/2}-1} \log %n^{-(1+2/\alpha_1-\alpha_1 \gamma_1/2)}. 
%$$ 

%%%%%%%%%%%%%%%%%%%%%%%%%%%%%%%%%%%%%%%%%%%%%%%%%%%%%%%%%%%%%%%%%%%%%%%%%%%%%%%%%%%%%%%%%%%%%%%%%%%%%%%%%%%%%%%%%%%%
%\noindent {\bf Proof of Lemma 2.} Since $X$ is continuously supported in $[a,b]$, it follows immediately that $R_0(c)$ and $R_1(c)$ are continuous in $c \in [a,b]$, and $R_0(C)$ and $R_1(c)$ are decreasing and increasing with respect to $c$, respectively.  Therefore, for any $\alpha$, the solution $c_\alpha^*$ to (\ref{eqn:neyman}) satisfies $R_0(c_\alpha^*)=P_{X|Y=-1}(X> c_\alpha^*) = \alpha$. 

%Let $w = 1-p(c_\alpha^*)$, then $c_\alpha^*=c_w^*$ by (\ref{eqn:cmt_w_sol}). The uniqueness follows from the strict monotonicity of $p(x)$ and Lemma 1. On the other hand, for any $w$, since $p(x)$ is strictly increasing, there is a unique solution $c_w^*$ to (\ref{eqn:cmt_w_sol}). Let $\alpha=P_{X|Y=-1}(X > c_w^*)$, then $c_\alpha^*=c_w^*$.

%%%%%%%%%%%%%%%%%%%%%%%%%%%%%%%%%%%%%%%%%%%%%%%%%%%%%%%%%%%%%%%%%%%%%%%%%%%%%%%%%%%%%%%%%%%%%%%%%%%%%%%%%%%%%%%%%%%%%%
\noindent {\bf Proof of Lemma 2.} For any given $z$, since $L_{\delta}(u)=L_{01} (u)+ \delta^{-1} (\delta-u)I(0 \leq u \leq \delta)$, we have
\begin{equation}
\begin{split}
E \Big ( L_{\delta}( Y(X-c(z)) )|Z=z \Big ) & =E \Big ( L_{01}( Y(X-c(z)) )|Z=z \Big ) \\
& + E\Big(\frac{\delta-Y(X-c(z))}{\delta} I(0 \leq Y(X-c(z)) \leq \delta) |Z=z \Big). \label{eqn:lemma2}
\end{split}
\end{equation}
Note that $E\big(\frac{\delta-Y(X-c(z))}{\delta} I(0 \leq Y(X-c(z)) \leq \delta) |Z=z \big)$ is decreasing in $\delta$, and approaches 0 when $\delta \rightarrow 0$. Furthermore, $E \big ( L_{01}( Y(X-c(z)) )|Z=z \big ) - E \big ( L_{01}( Y(X-c^*(z)) )|Z=z \big ) = \int_{c^*(z)}^{c(z)} (2 p_z (x)-1)f_z(x) dx$, which is increasing in $c(z)$ when $c(z)>c^*(z)$. Therefore, there exist $\delta_u (z)>0$ and $c_u (z)$ such that 
$$ 
\int_{c^*(z)}^{c_u(z)} (2 p_z (x)-1)f_z(x) dx \geq E\Big(\frac{\delta_u-Y(X-c)}{\delta_u} I(0 \leq Y(X-c) \leq \delta_u) |Z=z \Big).
$$
This implies that for any $\delta < \delta_u (z)$, $\argmin_{c} E \Big ( L_{\delta}( Y(X-c) )|Z=z \Big ) \leq c_u (z)$. Similarly, there exist $\delta_l(z)$ and $c_l (z)$ such that for any $\delta < \delta_l (z)$, $\argmin_{c} E \Big ( L_{\delta}( Y(X-c) )|Z=z \Big ) \geq c_l (z)$. Therefore, for any $\delta < \min \{\delta_l (z), \delta_u (z)\}$, $\argmin_{c} E \Big ( L_{\delta}( Y(X-c) )|Z=z \Big )$ must lie in a compact set ${\cal D}(z)$ around $c^*(z)$.

The second term on the right hand side of (\ref{eqn:lemma2}) is bounded below by 0 and above by $P\Big(|X-c| \leq \delta |Z=z \Big)$ and is decreasing in $\delta$. Therefore, by Dini's theorem, $P\Big(|X-c| \leq \delta |Z=z \Big)$ converges to 0 uniformly over ${\cal D} (z)$ as $\delta \rightarrow 0$. It further implies that $E \Big ( L_{\delta}( Y(X-c) ) |Z=z \Big)$ converges to $E \Big ( L_{01}( Y(X-c) ) |Z=z \Big)$ uniformly over ${\cal D} (z)$ as $\delta \rightarrow 0$. This, together with the fact that $E \Big ( L_{01}( Y(X-c) ) |Z=z \Big)$ is convex in $c$, implies that
$$
\argmin\limits_{c} E \Big ( L_{\delta}( Y(X-c(z)) )|Z=z \Big ) \longrightarrow \argmin\limits_{c} E \Big ( L_{01}( Y(X-c(z)) )|Z=z \Big ) = c^*(z),
$$
when $\delta$ converges to zero.

%%%%%%%%%%%%%%%%%%%%%%%%%%%%%%%%%%%%%%%%%%%%%%%%%%%%%%%%%%%%%%%%%%%%%%%%%%%%%%%%%%%%%%%%%%%%%%%%%%%%%%%%%%%%%%%%%%%%%%
Before delving into the proof of Theorem 2, we first define the $L_2$- metric entropy with bracketing for a function class ${\cal F}$. For any $\epsilon > 0$, $\{(l_1^l, l_1^u), \cdots, (l_m^l, l_m^u)\}$ forms an $\epsilon$-bracketing of $\cal F$, if for any $c \in {\cal F}$, there is a $j$, such that $l_j^l \leq l(c,\cdot) \leq l_j^u$ and $\max_{\{1\leq j \leq m\}} \|l_j^l-l_j^u\|_2 \leq \epsilon$, where $\| \cdot \|_2$ is the $L_2$-norm. Then the $L_2$-metric entropy of ${\cal F}$ with bracketing $H_B(\epsilon, {\cal F})$ is defined as a logarithm of the cardinality of the $\epsilon$-bracketing of $\cal F$ of the smallest size.

\noindent {\bf Proof of Theorem 2.} The proof consists of two steps. The first step establishes an upper bound for the misclassification error $e(\hat c, c^*)=E \big( L_{01}(Y(X-\hat c(Z))) - L_{01}(Y(X-c^*(Z))) \big)$, and the second step connects $e(\hat c, c^*)$ with $|\hat c(Z)-c^*(Z)|$ and attains the desired results.

{\it Step 1:} First we introduce some notations. Let $\tilde{l}_{\delta_n}(c,D_i)=L_{\delta_n} (y_i (x_i-c(z_i))) + \lambda J(c)$, where $D_i=(x_i, y_i, z_i)$. Similarly, denote $\tilde{l}(c,D_i)=L_{01} (y_i (x_i-c(z_i))) + \lambda J(c)$. Then the scaled empirical process $E_n (\tilde l(c,D)-\tilde l_{\delta_n}(c_0,D))$ is defined as
$$
E_n (\tilde l(c,D)-\tilde l_{\delta_n}(c_0,D) )=\frac{1}{n} \sum_{i=1}^n \Big ( \tilde{l}(c,D_i)-\tilde{l}_{\delta_n}(c_0,D_i)-E(\tilde{l}(c,D_i)-\tilde{l}_{\delta_n}(c_0,D_i)) \Big ).
$$
Since $L_{\delta_n} (y_i (x_i-c(z_i))) \geq L_{01} (y_i (x_i-c(z_i)))$ for any $\delta_n > 0$, we have
$$
\tilde{l}_{\delta_n}(c_0,D_i)-\tilde{l}(c,D_i) \geq \tilde{l}_{\delta_n}(c_0,D_i) - \tilde{l}_{\delta_n}(c,D_i).
$$
Furthermore, by Assumptions A-C,
\begin{eqnarray*}
e(c_0,c^*) &=& E L_{01}(Y(X- c_0(Z))) - E L_{01}(Y(X-c^*(Z))) \\
%&\leq& e_{\delta_n}(c_0,c^*)+P(|X-c^*(z)|\leq \delta_n|Z=z) \\
&\leq& e_{\delta_n}(c_0,c^*)+P \Big (|p_z(X)-p_z(c^*(z))|\leq a_4 \delta_n^{\gamma_2} \Big ) \leq  s_n + a_3 a_4^{\alpha_2} \delta_n^{\alpha_2 \gamma_2} \leq \beta_n^2/2.
\end{eqnarray*} 
%$$P(|X-c^*(z)|\leq \delta_n|Z=z)\leq P(|p_z(X)-p_z(c^*(z))|\leq a_4 \delta_n^{\gamma_2}) \leq a_3 a_4^{\alpha_2} \delta_n^{\alpha_2 \gamma_2}.$$
%Thus $e(c_0,c^*) \leq e_{\delta_n}(c_0,c^*)+a_3 a_4^{\alpha_2} \delta_n^{\alpha_2 \gamma_2}$. Furthermore, by Assumption A, it yields
%$$e(c_0,c^*) \leq s_n + a_3 a_4^{\alpha_2} \delta_n^{\alpha_2 \gamma_2} \leq \beta_n^2.$$
Let $\hat{c}=\argmin\limits_{c \in {\cal F}} \frac{1}{n}\sum\limits_{i=1}^n \tilde{l}_{\delta_n}(c,D_i)$ be the estimated personalized MCID, then
\begin{eqnarray*}
\{e(\hat{c},c^*) \geq \beta_n^2\} &\subset& \left \{\sup\limits_{ \{e(c,c^*) \geq \beta_n^2\}} \frac{1}{n} \sum_{i=1}^n \left( \tilde{l}_{\delta_n}(c_0,D_i) - \tilde{l}_{\delta_n}(c,D_i)\right) \geq 0 \right \}\\
&\subset& \left \{\sup_{\{e(c,c^*) \geq \beta_n^2\}} \frac{1}{n} \sum_{i=1}^n \left( \tilde{l}_{\delta_n}(c_0,D_i) - \tilde{l}(c,D_i)\right) \geq 0 \right \}.
\end{eqnarray*}
It immediately implies that 
$$
P(e(\hat{c},c^*) \geq \beta_n^2) \leq P^*\left( \sup_{\{ e(c,c^*) \geq \beta_n^2\}} \frac{1}{n} \sum_{i=1}^n \left( \tilde{l}_{\delta_n}(c_0,D_i) - \tilde{l}(c,D_i)\right) \geq 0 \right) \hat = I,
$$
where $P^*$ denotes the outer probability measure. 

Note that the functional space $\{c \in {\cal F}: e(c,c^*) \geq \beta_n^2\}$ can be partitioned as
$$
A_{ij}=\{c \in {\cal F}: 2^{i-1}  \beta_n^2 \leq e(c,c^*) < 2^i \beta_n^2, 2^{j-1} \max(J(c_0),1) \leq J(c) < 2^j \max(J(c_0),1) \};
$$
$$
A_{i0}=\{c \in {\cal F}: 2^{i-1} \beta_n^2 \leq e(c,c^*) < 2^i \beta_n^2, J(c) < \max(J(c_0),1) \},
$$
for $i=1,2,\cdots$ and $j=1,2,\cdots$. Then we need to establish some inequalities on the first and second moments of $\tilde l(c,D)- \tilde l_{\delta_n}(c_0,D)$ for $c \in A_{ij}$. 

For the first moment, note that for any $c \in {\cal F}$,
\begin{eqnarray*}
E(L_{01}(c,D)-L_{\delta_n}(c_0,D)) &=& E(L_{01}(c,D)-L_{01}(c^*,D))+ E(L_{01}(c^*,D)-L_{\delta_n}(c_0,D)) \\
&\geq& e(c,c^*) + e_{\delta_n}(c^*,c_0)-a_3 a_4^{\alpha_2}\delta_n^{\alpha_2 \gamma_2} \\
&\geq& e(c,c^*) -s_n-a_3 a_4^{\alpha_2}\delta_n^{\alpha_2 \gamma_2} 
\geq e(c,c^*) - \beta_n^2/2.
\end{eqnarray*}
Then with the assumption that $\lambda \max(J(c_0),1)\leq \beta_n^2/4$,
\begin{equation}
\inf\limits_{A_{ij}} E(\tilde{l}(c,D)-\tilde{l}_{\delta_n}(c_0,D)) \geq 2^{i-2} \beta_n^2 + (2^{j-1}-1)\lambda J(c_0)=M(i,j),
\label{eqn:entropy1}
\end{equation}
\begin{equation}
\inf\limits_{A_{i0}} E(\tilde{l}(c,D)-\tilde{l}_{\delta_n}(c_0,D)) \geq (2^{i-1}-3/4) \beta_n^2 \geq 2^{i-3} \beta_n^2 = M(i,0).
\label{eqn:entropy2}
\end{equation}
For the second moment, it follows from Assumptions B and C that for any $c \in {\cal F}$,
\begin{eqnarray*}
e(c,c^*)&=&E |p_Z(X)-1/2| \left|\sign(X-c^*(Z))-\sign(X-c(Z))\right|\\
        &\geq& \xi E\left|\sign(X-c^*(Z))-\sign(X-c(Z))\right| I(|p_Z(X)-1/2|\geq \xi)\\
        &\geq& \xi \left(E|\sign(X-c^*(Z))-\sign(X-c(Z))|-2 a_3 \xi^{\alpha_2} \right)\\
        &\geq& 2^{-1-2/\alpha_2}a_3^{-1/\alpha_2} \left(E|\sign(X-c^*(Z))-\sign(X-c(Z))|\right)^{(1+\alpha_2)/\alpha_2}\\
        & = & 2^{-1-2/\alpha_2}a_3^{-1/\alpha_2}\left(E|L_{01}(c^*,D)-L_{01}(c,D)|\right)^{(1+\alpha_2)/\alpha_2},
\end{eqnarray*}
with a choice of $\xi=(E|\sign(X-c^*(Z))-\sign(X-c(Z))|/4 a_{6})^{1/\alpha_2}$. Now we are ready to establish an upper bound for the second moment. Note that for any $d$, $L_{01}(c,D) \leq L_{\delta_n}(c,D)$, then $E(|L_{01}(c_0,D)-L_{\delta_n}(c_0,D)|)=E(L_{\delta_n}(c_0,D)-L_{01}(c_0,D)) = E(L_{\delta_n}(c_0,D)-L_{\delta_n}(c^*,D)+L_{\delta_n}(c^*,D)-L_{01}(c_0,D)) \leq e_{\delta_n} (c_0,c^*)+a_3 a_4^{\alpha_2} \delta_n^{\alpha_2 \gamma_2}$. Therefore, by the triangular inequality,
\begin{eqnarray*}
& & E\left(l(c,D)-l_{\delta_n}(c_0,D)\right)^2 \leq E(|l(c,D)-l_{\delta_n}(c_0,D)|) \\
&\leq& E|l(c^*,D)-l(c,D)| + E|l(c^*,D)-l(c_0,D)| + E|l(c_0,D)-l_{\delta_n}(c_0,D)|\\
&\leq& E|l(c^*,D)-l(c,D)| + E|l(c^*,D)-l(c_0,D)| + e_{\delta_n} (c_0,c^*)+a_3 a_4^{\alpha_2} \delta_n^{\alpha_2 \gamma_2}\\
&\leq& 2^{1+2/\alpha_2}a_3^{1/\alpha_2} (e(c,c^*)^{\alpha_2/(1+\alpha_2)}+e(c_0,c^*)^{\alpha_2/(1+\alpha_2)}) + e_{\delta_n} (c_0,c^*)+a_3 a_4^{\alpha_2} \delta_n^{\alpha_2 \gamma_2} \\
&\leq& a_6 (e(c,c^*))^{\alpha_2/(1+\alpha_2)},
\end{eqnarray*}
where $a_6=2^{2+2/\alpha_2}a_3^{1/\alpha_2}+1$, and the last inequality is due to the fact that $e(c,c^*) \geq \beta_n^2 \geq e_{\delta_n} (c_0,c^*)+a_3 a_4^{\alpha_2} \delta_n^{\alpha_2 \gamma_2} \geq e(c_0,c^*)$ for any $c \in A_{ij}$.
Consequently,
$$\sup\limits_{A_{ij}} E\left(l(c,D)-l_{\delta_n}(c_0,D)\right)^2 \leq v^2(i,j) \hat = 8 a_6 M(i,j)^{\alpha_2/(1+\alpha_2)},$$
where $i=1,2,\cdots$ and $j=0,1,2,\cdots$.

Now we are ready to establish the upper bound of $I$. Using (\ref{eqn:entropy1}) and (\ref{eqn:entropy2}), we have
\begin{eqnarray*}
I &\leq& \sum\limits_{i,j} P^*\left(\sup\limits_{A_{ij}}E_n(l_{\delta_n}(c_0,D)-l(c,D)) \geq M(i,j)\right)\\
  & & ~~~~~~+ \sum\limits_i P^*\left(\sup\limits_{A_{i0}}E_n(l_{\delta_n}(c_0,D)-l(c,D)) \geq M(i,0)\right) \hat = I_1 + I_2.
\end{eqnarray*}
Then we bound $I_1$ and $I_2$ separately by using Theorem 3 of Shen and Wong (1994), and we just need to verify the conditions (4.5)-(4.7) therein. To compute the metric entropy in (4.7), applying the same technique as in Shen et al. (2003) yields that $H_B(\epsilon, {\cal F}(2^j)) \leq H(\epsilon^2/2,{\cal G}(2^j))$ for any $\epsilon>0$ and $j=0,1,\cdots$, where ${\cal F}(2^j)=\{l(c,d)-l_{\delta_n}(c,d): c \in {\cal F}, J(c) \leq 2^j\}$. Since $\int^{v(i,j)}_{a_7 M(i,j)}H^{1/2}(u^2/2,{\cal G}(2^j))du/M(i,j)$ is non-increasing in $i$ and $M(i,j)$, we have
\begin{eqnarray*}
& & \int^{v(i,j)}_{a_7M(i,j)}H^{1/2}(u^2/2,{\cal G}(2^j))du/M(i,j) \\
&\leq& \int^{(8a_6)^{1/2}M(1,j)^{\alpha_2/2(\alpha_2+1)}}_{a_7M(1,j)} H^{1/2}(u^2/2,{\cal G}(2^j))du/M(1,j) \leq \phi(\epsilon_n,2^j),
\end{eqnarray*}
where $a_7=1/64$. Simply let $\epsilon=1/2$, then Assumption D implies (4.7). Furthermore, (4.5) and (4.6) are satisfied with the above choice of $\epsilon, M(i,j), v(i,j)$ and $T=1$. In more details, (4.7) implies (4.5) and $M(i,j)/v^2(i,j) \leq 1/8$ implies (4.6).

Then Theorem 3 of Shen and Wong (1994) with $M=n^{1/2}M(i,j),v=v^2(i,j), \epsilon=1/2$ and $T=1$ implies that 
\begin{eqnarray*}
I_1 &\leq& \sum\limits_{j=1}^{+\infty} \sum\limits_{i=1}^{+\infty} 3 \ \exp\left(-\frac{n M(i,j)^2}{4(4v^2(i,j)+ M(i,j)/3)}\right) \leq  \sum\limits_{j=1}^{+\infty} \sum\limits_{i=1}^{+\infty} 3 \ \exp\left(-a_8 n M(i,j)^{\frac{\alpha_2+2}{\alpha_2+1}}\right)\\
&\leq& \sum\limits_{j=1}^{+\infty} \sum\limits_{i=1}^{+\infty} 3 \  \exp\left(-a_8 n[2^{i-2}\beta_n^2+(2^{j-1}-1)\lambda J(c_0)]^{\frac{\alpha_2+2}{\alpha_2+1}} \right)\\
%&\leq& \sum\limits_{j=1}^{+\infty} \sum\limits_{i=1}^{+\infty} 3 \  \exp\left(-a_8 n[(2^{i-2}\beta_n^2)^{\frac{\alpha_2+2}{\alpha_2+1}}+(2^{j-1}-1)\lambda J(c_0)^{\frac{\alpha_2+2}{\alpha_2+1}}] \right)\\
&\leq&  3 \exp\left(-a_8 n (\lambda J(c_0))^{\frac{\alpha_2+2}{\alpha_2+1}}\right) \left(1-\exp(-a_8 n (\lambda J(c_0))^{\frac{\alpha_2+2}{\alpha_2+1}})\right)^{-2},
\end{eqnarray*}
where $a_8$ is a positive constant. $I_2$ can be bounded similarly, and thus
$$
I \leq  6 \exp\left(-a_8 n (\lambda J(c_0))^{\frac{\alpha_2+2}{\alpha_2+1}}\right) \left(1-\exp(-a_8 n (\lambda J(c_0))^{\frac{\alpha_2+2}{\alpha_2+1}})\right)^{-2},
$$
which implies that $I^{1/2} \leq (2.5+I^{1/2}) \exp \left(-a_8 n (\lambda J(c_0))^{\frac{\alpha_2+2}{\alpha_2+1}}\right)$. With $I \leq I^{1/2} \leq 1$, then 
\begin{equation}
P \Big (e(\hat c,c^*) \geq \beta_n^2 \Big ) \leq 3.5 \exp \Big ( -a_{8} n (\lambda_n J(c_0))^{\frac{\alpha_2+2}{\alpha_2+1}} \Big ) .
\label{eqn:err_bound}
\end{equation}

{\it Step 2:} For any given $z \in {\cal D}_Z$, Assumptions B and C are similar to (\ref{eqn:low_noise}) and (\ref{eqn:lipschitz}) in Theorem 1 and yield that for any sufficiently small $\xi > 0$, there exists a constant $a_9 > 0$ such that 
$$
E \Big( L_{01}( Y(X-(c^*(z)\pm \xi) )|Z=z \Big)- E \Big( L_{01}( Y(X-c^*(z)) )|Z=z \Big) \geq a_9 \xi^{1+2/\alpha_2},
$$
and $E \Big( L_{01}( Y(X-(c^*(z)\pm \xi) )|Z=z \Big)- E \Big( L_{01}( Y(X-c^*(z)) )|Z=z \Big)$ is monotonically increasing with $\xi$. Therefore, conditional on the training data,
\begin{eqnarray*}
e(\hat{c},c^*) & \geq & E \Big( \{L_{01}( Y(X- \hat{c}(Z)))-  L_{01}( Y(X-c^*(z)) )\}  I(|\hat{c}(Z)-c^*(Z)| \geq \xi) \Big)\\
& \geq &  E \Big( \{L_{01}( Y(X- (c^*(Z)+\xi \sign(\hat{c}(Z)-c^*(Z)))))  \\
&  & \qquad \qquad \qquad \qquad \qquad - L_{01}( Y(X-c^*(z)) )\}  I(|\hat{c}(Z)-c^*(Z)| \geq \xi) \Big)\\
& \geq & a_9 \xi^{1+2/\alpha_2} E(I(|\hat{c}(Z)-c^*(Z)| \geq \xi)) = a_9 \xi^{1+2/\alpha_2} P(|\hat{c}(Z)-c^*(Z)| \geq \xi).
\end{eqnarray*}

We are now ready to bound $P \Big (|\hat c(Z)-c^*(Z)| \geq (\beta_n^2 \log(1/\beta_n^2) )^{\frac{ \alpha_2}{\alpha_2+2}} \Big )$. By (\ref{eqn:err_bound}) and the above inequality,
\begin{equation*}
\begin{split} 
P \Big ( & |\hat c(Z)-c^*(Z)| \geq (\beta_n^2 \log(1/\beta_n^2) )^{\frac{ \alpha_2}{\alpha_2+2}}  \Big ) = E \Big ( P \big ( |\hat c(Z)-c^*(Z)| \geq (\beta_n^2 \log(1/\beta_n^2) )^{\frac{ \alpha_2}{\alpha_2+2}} \big ) \Big ) \\
& \leq E(a_9^{-1} \beta_n^{-2} \log(1/\beta_n^2)^{-1} e(\hat{c},c^*) )  \leq  3.5 \exp \Big ( -a_{8} n (\lambda_n J(c_0))^{\frac{\alpha_2+2}{\alpha_2+1}} \Big ) + a_9^{-1} (\log(1/\beta_n^2))^{-1}.
\end{split}
\end{equation*}
The desired results follow immediately after re-defining $a_9$.

%which, together with the first part, implies $P(|\hat{c}(Z)-c^*(Z)| \geq \xi) = O_p (\beta_n^2 \xi^{-(1+2/\alpha_2)})$. The desired result follows immediately.

%Next, denote $s_n (c) = n^{-1} \sum_{i=1}^n L_\delta \Big(y_i (x_i-c(z_i))\Big) + \lambda w^T K w/2$ and $\hat{c} (z) = \hat{b} + \hat{w}^T K(z)$, where $K(z) = (K(z_1, z),\ldots, K(z_n, z))$. Since $\hat{c}$ is the minimizer of $s_n (c)$, $\lambda \hat{w}^T K \hat{w}/2 \leq s_n (\hat{c}) \leq s_n (0)$. Further,  $ \lambda \hat{w}^T K \hat{w} \geq \lambda_{K,min}  \hat{w}^T \hat{w}$, where $\lambda_{K,min}$ is the smallest eigenvalue of $K$ and thus $\| \hat w\|_2^2 \leq 2 s_n(0) / (\lambda \lambda_{K,min})$. Moreover, by the assumption that $Z$ is defined on a compact set $D_Z$, we have $\|K(z)\|_2$ and $b$ are bounded. By Cauchy-Schwarz inequality, $\hat{c} (z) \leq \|\hat w\|_2 \|K(z)\|_2 + b$. Therefore, by $|c^*(z)| \leq B$, $|\hat{c}(z) - c^*(z)|$ is uniformly bounded. Then by using the fact that $E(|\hat{c}(Z)-c^*(Z)|) = E(|\hat{c}(Z)-c^*(Z)| I(|\hat{c}(Z)-c^*(Z)| < \xi)) + E(|\hat{c}(Z)-c^*(Z)| I(|\hat{c}(Z)-c^*(Z)| \geq \xi))$, it follows immediately that $E(|\hat{c}(Z)-c^*(Z)|) = O_p (\beta_n^{\alpha_2/(1+\alpha_2)})$ when $\xi = \beta_n^{\alpha_2/(1+\alpha_2)}$.  

%%%%%%%%%%%%%%%%%%%%%%%%%%%%%%%%%%%%%%%%%%%%%%%%%%%%%%%%%%%%%%%%%%%%%%%%%%%%%%%%%%%%%%%%%%%%%%%%%
\section*{Appendix B: a counterexample for various losses}

Consider a simple example where $X$ is uniformly distributed on $[a,b]$, $p(x)$ is continuous and strictly increasing, and $\min\{c^*-a, b-c^*\}>1$. By (\ref{eqn:cmt_sol}), $c^*$ is the unique MCID. On the other hand, the minimizer of the hinge loss must satisfy that
$$
\int_{a}^{c^*+1} p(x) dx - \int_{c^*-1}^b (1-p(x)) dx =0,
$$
the minimizer of the logistic loss must satisfy that
$$
\int_{a}^b \Big ( p(x)- \frac{1}{1+e^{c^*-x}} \Big ) dx=0,
$$
and the minimizer of  the $\psi$-loss must satisfy that 
$$
\int_{c^*-1}^{c^*+1} \Big ( p(x)-\frac{1}{2} \Big ) dx=0.
$$
These equalities do not hold in general.  For instance, let $p''(x)=0$, when $x \geq c^*$ and $p''(x)>0$ otherwise, then minimizers for all three losses are strictly larger than $c^*$.

{}

\newpage

\begin{table}[!ht]
      \begin{center}
      \caption{Simulation I. Averaged MCID and the misclassification error (MCE) and their standard errors (in parentheses) for our method (OUR) and the method by Shiu and Gatsonis (SG) based on 100 replications. The ideal performance is included as the baseline for comparison. }
      \label{tab:sim1}
      \medskip
        \begin{tabular}{cccccc}
        \hline
         \hline
              & &~n=250~~ & ~n=500~~ & ~n=1000~ & ~Ideal~  \\
              \hline
             \multicolumn{6}{c}{\it Example 1} \\
             \hline
              & OUR & 0.055(0.0116)  & -0.021(0.0058)& 0.004(0.0032)  &  \\
              [-1ex] \raisebox{1.5ex}{MCID} & SG & 0.078(0.0387) & -0.065(0.0290) & -0.080(0.0222) &\raisebox{1.5ex}{0.000}\\
              \hline
              & OUR & 0.260(0.0010)  &0.255(0.0005)   &0.253(0.0003)  & \\
              [-1ex] \raisebox{1.5ex}{MCE} & SG &0.344(0.0045)  &0.355(0.0033)   &0.374(0.0024)  &\raisebox{1.5ex}{0.250}
                \\
              \hline
              \multicolumn{6}{c}{\it Example 2} \\
              \hline
              & OUR & -0.563(0.0187) &-0.496(0.0095) &-0.497(0.0056)  & \\
              [-1ex] \raisebox{1.5ex}{MCID} & SG & -0.436(0.0827)  & -0.286(0.0676) &-0.370(0.0526)  &\raisebox{1.5ex}{-0.514} \\
              \hline
              & OUR & 0.257(0.0009)  &0.253(0.0005)   & 0.252(0.0003)  &   \\
              [-1ex] \raisebox{1.5ex}{MCE} & SG & 0.338(0.0043) &0.361(0.0033)   &0.374(0.0024)  &\raisebox{1.5ex}{0.250}  \\
              \hline
               \hline
        \end{tabular}
        \label{tab:population-based_sim}
       \end{center}
\end{table}

\begin{table}[!ht]
      \begin{center}
      \caption{Simulation II. Estimated means and standard deviations (in parentheses) of the misclassification error by using our proposed method with linear and Gaussian kernels based on 50 replications.}
        \begin{tabular}{ccccc}
         \hline
         \hline
                    &~n=100~ & ~n=250~ & ~n=500~ & 
                    ~Ideal~  \\
              \hline
              & \multicolumn{3}{c}{\it Example 1} & \\
              \hline
              Linear & 0.256(0.0119)& 0.254(0.0112)& 0.250(0.0108)  &\\
              [-1ex] Gaussian & 0.280(0.0177) & 0.270(0.0146) & 0.259(0.0130) &\raisebox{1.5ex}{0.250}\\
              \hline
              & \multicolumn{3}{c}{\it Example 2} & \\
              \hline
              Linear & 0.412(0.0146) & 0.408(0.0140) & 0.408(0.0095)   &\\
              [-1ex] Gaussian & 0.290(0.0169) & 0.274(0.0133)   & 0.260(0.0118) &\raisebox{1.5ex}{0.250}\\   
             \hline
              & \multicolumn{3}{c}{\it Example 3} & \\
              \hline
              Linear  & 0.315(0.0132) & 0.313(0.0129)& 0.318(0.0103)  &\\
              [-1ex] Gaussian & 0.323(0.0182)  & 0.308(0.0122) &0.293(0.0109)  &\raisebox{1.5ex}{0.250}\\
          \hline
        \hline
        \end{tabular}
        \label{table2}
       \end{center}
\end{table}

\begin{table}[!ht]
      \begin{center}
      \caption{Real applications. Averaged MCID and misclassification error (MCE) and their standard errors(in parenthesis) by using the method by Shiu and Gatsonis (SG), the population-based MCID (OUR), the personalized MCID with linear kernel (OUR$_L$) and Gaussian kernel (OUR$_G$) based on 50 replications.}
       \medskip
        \begin{tabular}{ccccc}
        \hline
         \hline
              &~SG~&~OUR~&~OUR$_L$~&~OUR$_G$~  \\         
             \hline
             \multicolumn{5}{c}{\it WHMBL } \\
                \hline
               MCID & -45.004(3.3011)& 20.610(0.4905)&~~-~~ &~~-~~ \\
               MCE& 0.436(0.0016) & 0.358(0.0014) & 0.365(0.0186) &0.376(0.0185)\\            
              \hline
               \multicolumn{5}{c}{\it Hot Flush} \\
               \hline
               MCID & 5.426(0.4453) & 6.060(0.0229) &~~-~~ &~~-~~ \\
               MCE & 0.399(0.0049) & 0.282(0.0005) & 0.260(0.0054)& 0.268(0.0031)\\  
             \hline
               \hline
        \end{tabular}
        \label{table4}
       \end{center}
\end{table}

%\begin{figure}[!h]
%\caption{The estimated MCID with linear kernel $\hat{c}_L$ and with Gaussian kernel $\hat{c}_G$ in a randomly selected replication of Example 1 when $n=250$ and $Z_2=0$.}
%\begin{center}
%\includegraphics[width=0.75\textwidth]{EstC.pdf}
%\label{fig:EstC}
%\end{center}
%\end{figure}

\begin{figure}[!h]
\caption{Sensitivity analysis of $\delta$ in a randomly selected replication of Example 1 with $n=250$.}
\begin{center}
\includegraphics[width=0.75\textwidth]{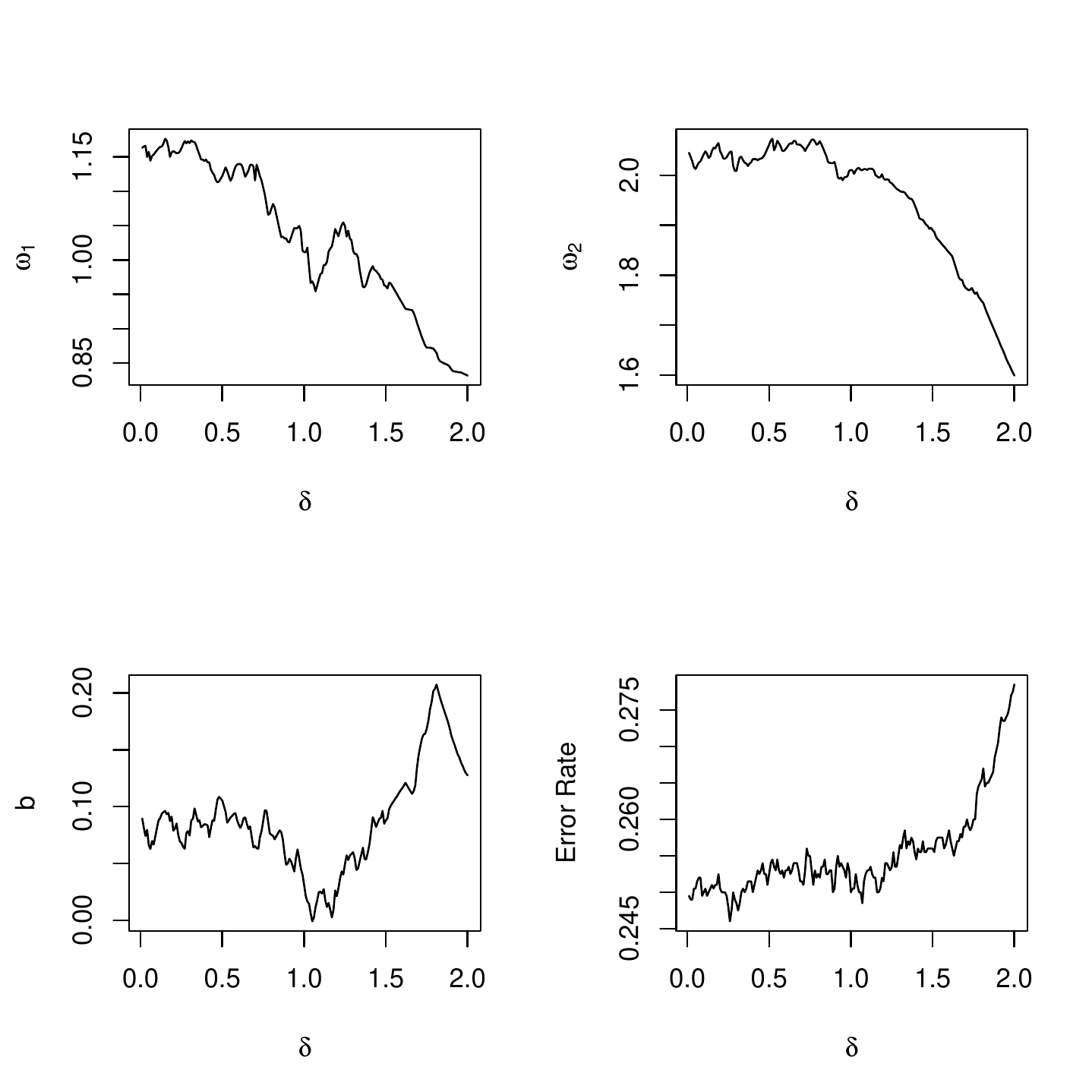}
\label{fig:Sensitivity}
\end{center}
\end{figure}

\end{document}